\documentclass[iop]{emulateapj}

\def\na{New A}
\def\icarus{Icarus}
\def\sun{\odot}

\usepackage{epsfig}
\begin{document}
\title{Runaway Growth During Planet Formation: Explaining the Size
Distribution of Large Kuiper Belt Objects} \shortauthors{Schlichting \& Sari}
\shorttitle{The Kuiper Belt Size Distribution} \author{Hilke
E. Schlichting\altaffilmark{1,2,3} and Re'em Sari\altaffilmark{4,5}
}\altaffiltext{1} {UCLA, Department of Earth and Space Science, 595 Charles E.
Young Drive East, Los Angeles, CA 90095}\altaffiltext{2} {CITA, University of
Toronto, 60 St. George St., ON, M5S 3H8, Canada} \altaffiltext{3} {Hubble
Fellow}\altaffiltext{4}{Racah Institute of Physics, Hebrew University,
Jerusalem 91904, Israel} \altaffiltext{5}{California Institute of Technology,
MC 130-33, Pasadena, CA 91125} \email{hilke@ucla.edu}

\begin{abstract} 
Runway growth is an important stage in planet formation during which large
protoplanets form, while most of the initial mass remains in small
planetesimals. The amount of mass converted into large protoplanets and their
resulting size distribution are not well understood. Here, we use analytic
work, that we confirm by coagulation simulations, to describe runaway growth
and the corresponding evolution of the velocity dispersion. We find that
runaway growth proceeds as follows: Initially all the mass resides in small
planetesimals, with mass surface density $\sigma$, and large protoplanets
start to form by accreting small planetesimals. This growth continues until
growth by merging large protoplanets becomes comparable to growth by
planetesimal accretion. This condition sets in when $\Sigma/\sigma \sim
\alpha^{3/4} \sim 10^{-3}$, where $\Sigma$ is the mass surface density in
protoplanets in a given logarithmic mass interval and $\alpha$ is the ratio of
the size of a body to its Hill radius. From then on, protoplanetary growth and
the evolution of the velocity dispersion become self-similar and $\Sigma$
remains roughly constant, since an increase in $\Sigma$ by accretion of small
planetesimals is balanced by a decrease due to merging with large
protoplanets. We show that this growth leads to a protoplanet size
distribution given by $N(>R) \propto R^{-3}$ where $N(>R)$ is the number of
objects with radii greater than $R$ (i.e., a differential power-law index of
4). Since only the largest bodies grow significantly during runaway
growth, $\Sigma$ and thereby the size distribution is preserved.  We apply our
results to the Kuiper Belt, which is a relic of runaway growth where planet
formation never proceeded to completion. Our results successfully match the
observed Kuiper belt size distribution, they illuminate the physical processes
that shaped it and explain the total mass that is present in large Kuiper belt
objects (KBOs) today. This work suggests that the current mass in large KBOs
is primordial and that it has not been significantly depleted. We also predict
a maximum mass-ratio of Kuiper belt binaries that formed by dynamical
processes of $\alpha^{-1/4} \sim 10$, which explains the observed clustering
in binary companion sizes that is seen in the cold classical belt. Finally,
our results also apply to growth in debris disks, as long as frequent
planetesimal-planetesimal collisions are not important during the growth.
\end{abstract}


\section{INTRODUCTION}
Ice giants, the cores of gas giants and protoplanets that later form
terrestial planets, are generally believed to have formed by coagulation from
small planetesimals. Understanding the evolution of the size distribution of
growing protoplanets, their velocity dispersion and the interplay between the
two is crucial for shedding light onto the planet formation process. Here, we
study the runaway growth of protoplanets \citep[e.g.][]{S72,G78,WS89,KI96} and
their subsequent velocity evolution. Runaway growth can occur when the
accretion cross section of protoplanets is enhanced by gravitational
focusing. Since, gravitational focusing is strongest for the largest bodies,
the radii of larger protoplanets run away from that of smaller. This leads to
a size distribution that develops a tail containing a small number of large
protoplanets. It is however not always clear what fraction of the total mass
participates in this runaway growth and what protoplanet size distribution
such runaway growth gives rise to \citep{MHL00,MG01}. In this paper we address
both of these questions. In the following sections, we discuss runaway growth
in the context of the Kuiper belt, which is an ideal laboratory to test our
results, since it is a remnant of the primordial Solar system, where planet
formation never reached completion. The results however, also apply to runaway
growth during planet formation and growth in debris disk, as long as gas plays
no significant role in the accretion and damping of the velocity dispersion.

The Kuiper belt consists of a disk of icy bodies located at the outskirts of
our planetary system, just beyond the orbit of Neptune and contains some of
the least processed bodies in our Solar system. Motivated by the discovery of
the first Kuiper belt object \citep{JL93} after Pluto and Charon, several
groups conducted large scale surveys to characterize the Kuiper belt. These
efforts led to the discovery of more than 1200 objects in the Kuiper belt to
date. The Kuiper belt size distribution contains many important clues
concerning the formation of Kuiper belt objects (KBOs), their effective
strength and their collisional evolution \citep{D69,SC97,DF97,KL99,PS05}. It
also provides a snapshot of an earlier stage of planet formation, which was
erased elsewhere in the Solar system where planet formation proceeded all the
way to completion. The cumulative size distribution of KBOs larger than $R
\gtrsim 50~\rm{km}$ is well described by a single power-law given by
\begin{equation}\label{e1}
N(>R) \propto R^{1-q}
\end{equation} 
where $N(>R)$ is the number of objects with radii greater than $R$, and $q$ is
the power-law index. Kuiper belt surveys find that the size distribution for
KBOs with radii greater than about 50~km follows this power-law with $q \sim
4$ \citep[e.g.][]{TJL01,BTA04,FH108,FK108}, which implies roughly equal mass
per logarithmic mass interval. This size distribution is a relic of the
accretion history in the Kuiper belt and therefore provides valuable insights
into the formation of large KBOs ($R \gtrsim 50~\rm{km}$)
\citep[e.g.][]{S96,DF97,K02}. Observations suggest that there is a break in
the power-law size distribution at smaller KBO sizes
\citep[e.g.][]{BTA04,FH108,FK108,SO09,FH10}. The break in the size
distribution is generally attributed to collisions that break-up small KBOs
(i.e.,$R \lesssim 50~\rm{km}$ ) and modify their size distribution
\citep[e.g.][]{D69,KB04,PS05}. The KBO size distribution below the break is
still poorly constrained, although some encouraging progress has been made
recently in probing the abundance of sub-km-sized KBOs by stellar occultations
\citep[e.g.][]{LC08,SO09,B10}.

The work presented in this paper focuses on the size distribution of large KBOs
($R \gtrsim 50~\rm{km}$), which is well constrained by observations and which
sheds light onto the formation of KBOs, protoplanets and accretion processes
that could be ongoing in other debris disks. Numerical coagulation simulations
have been successful in reproducing the observed KBO size distribution. Such
simulations typically find that the accretion processes of KBOs yield a
power-law size distribution with $q \sim 3.8 - 4.5$ for 10-100~km and larger
objects \citep{KL99,K02,KB04}, which is consistent with the observed power-law
size distribution. Despite their success, the reason for the actual slope of
the distribution has so far not been explained by such simulations. In this
paper, we offer an explanation for the slope of the KBO size distribution and
for the amount of mass in the large KBOs that are observed in today's Kuiper
belt. Specifically we find a power-law index of $q \sim 4$ and a total mass in
large KBOs of $\sim 10^{-3}$ of the initial planetesimal mass, which is
consistent with the current observed mass in the Kuiper belt. We also make a
prediction for the maximum mass ratio of Kuiper belt binaries that formed by
dynamical processes, for example, by three body interactions, and show
that our prediction is in good agreement with the observations. Although our
work focuses on the Kuiper belt, the results also apply to early stages of
planet formation and protoplanetary growth in debris disks.

Our paper is structured as follows: In \S 2 we analytically describe the
growth of large KBOs, including their velocity dispersion and derive the slope
of the KBO size distribution. We confirm our analytic results in \S 3 with
coagulation simulations. In \S 4, we discuss how semi-collisional accretion,
binary mergers and frequent planetesimal collisions would affect our
results. We show that our results on the KBO growth and velocity dispersion
have interesting implications for the formation of Kuiper belt binaries and
predict the maximum mass ratio for binaries that formed by dynamical processes
in \S 5. Discussion and conclusions follow in \S 6.

\section{Analytical Treatment}\label{s1}
In order to gain an analytic understanding of the growth processes of large
KBOs and the associated velocity evolution we use the 'two-groups
approximation' \citep{GLS02,GLS04}. The `two-groups approximation' consists of
the identification of two groups of objects, small ones, that contain most of
the total mass with mass surface density $\sigma$, and large ones, that
contain only a small fraction of the total mass with mass surface density
$\Sigma \ll \sigma $. We define $\Sigma$ as the mass surface density in a
single logarithmic mass interval, that includes the largest bodies formed at a
given time. In contrast $\sigma$ is defined as the total mass in small
objects.

Within the framework of the `two-groups approximation' we arrive at the
following picture for KBO growth. Initially all the mass is in small
bodies. As the small bodies start to accrete each other, large bodies begin to
form. To simplify the argument, we only consider the mass surface density of
the small and large bodies here, ignoring intermediate size bodies for now. As
we show later, the large and small bodies alone determine the velocity
dispersion for bodies of all sizes and only the large bodies grow
significantly. In the initial stage, $\Sigma$ grows due to the accretion of
small bodies. Therefore the size of the largest bodies and the total mass in
large bodies increases with time. During this growth phase the velocity
dispersion of the small bodies increases due to viscous stirring by the
large bodies. The velocity dispersion of large bodies is damped by
dynamical friction provided by the small bodies. $\Sigma$ continues to grow
until the growth of large KBOs by accretion of comparable size objects starts
to compete with growth by accretion of small bodies. From then on, $\Sigma$
remains roughly constant in a given logarithmic mass interval, while the size
of the large KBOs grows linearly with time. How the KBO growth ended and
how exactly the small bodies were lost from the Kuiper belt are still the
subject of ongoing research and are unimportant for the purpose of this paper
and we therefore will not discuss them here further. We confirm the outlined
KBO growth analytically and with numerical simulations. We show that the
mass-ratio, $\Sigma/\sigma$, is not arbitrary but an outcome of KBO
growth.

\subsection{Growth and Velocity evolution}
Large KBOs viscously stir the small bodies, increasing the small bodies'
velocity dispersion $u$. As a result $u$ grows on the same timescale as $R$,
as long as the small bodies experience no significant damping by either gas or
mutual collisions, which are, most likely, not yet important (see section
\ref{s3}). We can therefore write the evolution of the small bodies' velocity
dispersion as
\begin{equation}\label{e02}
\frac{1}{u}\frac{du}{dt} \sim \Omega \frac{\Sigma}{\rho R}
\alpha^{-2}\left(\frac{u}{v_H}\right)^{-4}
\end{equation}
where $\Omega$ is the orbital frequency around the Sun, $R$ the radius of the
the large bodies and $\rho$ is their material density. The Hill velocity of
the large bodies is $v_H$, which is given by $v_H = \Omega R_H$ where $R_H$ is
the Hill radius, $R_H=a(M/3M_{\sun})^{1/3}$, where $M_{\sun}$ is the mass of
the Sun, $M$ and $a$ are the mass and semi-major-axis of the large
KBOs. Finally, $\alpha = R/R_{H}$ and is $\sim 10^{-4}$ at the distance of the
Kuiper belt. We assumed in writing the expression for $u$ that $u > v_H$, we
will verify that this is the correct and self-consistent velocity regime for
$u$ at the end of this section. Initially, the large bodies grow by accreting
the small ones. Their growth rate is given by
\begin{equation}\label{e03}
\frac{1}{R}\frac{dR}{dt} \sim \Omega \frac{\sigma}{\rho R} \alpha^{-1}
\left(\frac{u}{v_H}\right)^{-2}
\end{equation}
where we assumed that $v<v_H$, which we verify later in this section.
Equating the rates from equations (\ref{e02}) and (\ref{e03}) and solving for
$u$ we find
\begin{equation}\label{e2}
\frac{u}{v_H} \sim \left( \frac{\Sigma}{\sigma \alpha} \right)^{1/2}
\end{equation}
\citep{GLS04}. The velocity $v$ of large
KBOs increases due to mutual viscous stirring, but is damped by dynamical
friction from the sea of small bodies such that $v < u$. The competition
between the stirring and damping can be written as 
\begin{equation}\label{e2_3}
\frac{1}{v}\frac{dv}{dt} \sim \Omega \frac{\Sigma}{\rho R} \alpha^{-2}
\left(\frac{v}{v_H}\right)^{-1}-\Omega \frac{\sigma}{\rho R} \alpha^{-2}
\left(\frac{u}{v_H}\right)^{-4}.
\end{equation}
Balancing the stirring and damping rates for $v$ and substituting for $u$ from
equation (\ref{e2}) into equation (\ref{e2_3}), we find
\begin{equation}\label{e3}
\frac{v}{v_H} \sim \alpha^{-2} \left( \frac{\Sigma}{\sigma}\right)^{3}.
\end{equation}
Having derived expressions for the large and small bodies' velocity dispersions
we now turn to examining the growth in more detail.

Expressions similar to the ones above have been widely used in the literature
\citep[e.g.][]{S72,G91,DT93,R2003,GLS04,CLM07}. However, in these works
$\Sigma/\sigma$ was treated as a free parameter. Here we present our method
for deriving $\Sigma/\sigma$ and calculate its value during runaway growth.

Within our `two-groups approximation' large bodies have two distinct modes for
growth. In the first, they can grow by the accretion of small bodies. In the
second, they grow by accreting objects comparable to their own
size. The growth rate for large KBOs is given by
\begin{equation}\label{e4}
\frac{1}{R} \frac {dR}{dt} \sim \Omega \frac{\sigma}{\rho R} \alpha^{-1}
\left(\frac{u}{v_H}\right)^{-2} + \Omega \frac{\Sigma}{\rho R} \alpha^{-3/2} 
\end{equation}
where we used the accretion rate corresponding to sub-Hill velocity
dispersions for large bodies and the corresponding rate for $u>v_H$ for small
bodies. The first term in expression (\ref{e4}) describes the growth of large
KBOs by the accretion of small bodies and the second corresponds to the growth
of large KBOs by accreting objects of their own size. Substituting the
expression for $u$ from equation (\ref{e2}) we have
\begin{equation}\label{e5}
\frac{1}{R} \frac {dR}{dt} \sim \Omega \frac{\sigma}{\rho R}
\left(\frac{\sigma}{\Sigma}\right) + \Omega \frac{\Sigma}{\rho R}
\alpha^{-3/2}.
\end{equation}
Comparing the two terms in expression (\ref{e5}) we find that they contribute
about equally to the growth of large KBOs, if
\begin{equation}\label{e6}
\frac{\Sigma}{\sigma} \sim \alpha^{3/4} \sim 10^{-3}.
\end{equation} 
Therefore, growth of large objects will be dominated by the accretion of small
bodies, if $\Sigma/\sigma \ll \alpha^{3/4}$. If on the other hand
$\Sigma/\sigma \gg \alpha^{3/4}$, then accretion of comparable size objects
will be the dominant mode of growth provided that $v$ remains less than the
Hill velocity. Since initially $\Sigma/\sigma \ll \alpha^{3/4}$ it follows
that the growth of large KBOs was at the beginning dominated by the accretion
of small bodies, as assumed in equation (\ref{e03}). This mode of growth
continues until $\Sigma/\sigma \sim \alpha^{3/4}$ at which stage accretion of
comparable size bodies starts to compete with growth by accretion of small
bodies. From then on, large KBOs grow in roughly equal amounts by accreting
small bodies and by merging with comparably-sized KBOs. As a result $\Sigma$
remains constant, since its increase due to accretion of small bodies is
counteracted by a decrease due to the accretion of large KBOs. The radii of
large KBOs grows linearly with time (see equation [\ref{e5}]). We find it very
encouraging that the observed mass in large KBOs in each logarithmic size
interval estimated from recent Kuiper belt surveys
\citep[e.g.][]{PK08,TB03,TJL01} is about $10^{-3}$ of that of the minimum mass
solar nebula (MMSN) \citep{H81} extrapolated to a heliocentric distance of
40~AU. The observed mass in large KBOs ($R \gtrsim 50~\rm{km}$) is therefore
consistent with $\Sigma/\sigma \sim 10^{-3}$ and with the hypothesis that they
formed by coagulation in the Kuiper belt from an MMSN. The growth of KBOs ended
in this runaway phase with $\Sigma \sim \alpha^{3/4} \sigma \sim 10^{-3}
\sigma$. In the context of planet formation this phase of runaway growth is
terminated by the onset of oligarchic growth, when each large body dominates
the stirring in its own feeding zone, allowing $\Sigma$ to become comparable
to $\sigma$.

Given this understanding of the KBO growth we now show that our choices for
the velocity regimes of $u$ and $v$ are self-consistent with this
picture. Substituting $\Sigma/\sigma \sim 10^{-3}$ into equation (\ref{e2}),
we find that $u \sim 3 v_H$. This implies that $u$ is about a few times the
Hill velocity of the large bodies. We note here that our derivation of the
expression for $u$, specifically equating equations (\ref{e03}) and
(\ref{e02}), remains valid even when the large bodies contribute significantly
to the growth of large KBOs, because the growth rates due to the accretion of
small and large bodies are comparable. Evaluating equation (\ref{e3}) we find
that $v \sim 0.1 v_H$. This implies that the velocity dispersion of large
bodies was sub-Hill (i.e. $v<v_H$) during the formation of large 100-km-sized
KBOs while that of the small bodies was super-Hill (i.e. $u>v_H$). This
confirms that we used the correct velocity regime for $u$ and $v$ in the
derivation of the KBO accretion rates above, ensuring that our treatment is
self-consistent.

Rewriting equation (\ref{e6}) as 
\begin{equation}\label{e7}
\Sigma \sim \sigma \alpha^{3/4} \sim \textrm{constant}.
\end{equation}
As we have shown above, $\Sigma$ of the largest bodies is constant with a
value of $\sim \alpha^{3/4} \sigma$. Once large bodies form, their mass per
logarithmic mass interval is, apart from a very brief period (see section
\ref{subHill2}), preserved. This is because bodies smaller than the largest
objects, do not grow significantly on the growth timescale of the largest
objects, i.e. the growth is in the runaway regime (see section
\ref{superHill}). Furthermore, such smaller bodies are also not efficiently
consumed by larger objects. Their mass surface density and size
distribution are therefore frozen (see section \ref{superHill}). Thus,
$\Sigma$ is constant in time, which results in 
\begin{equation}\label{e8}
N(>R) \propto R^{-3},
\end{equation}
since $\Sigma \propto N(>R) R^3$. This implies a power-law index $q=4$ (see
equation (\ref{e1})) for large KBOs.

Our work suggests that the growth of large KBOs resulted from the accretion of
small and large KBOs alike and that this mode of growth in the runaway regime
gave rise to the observed size distribution of large KBOs. We confirm this
result in section \ref{s2} using coagulation simulations. Our findings are
consistent with direct observations of the size distribution of large KBOs
\citep{BTA04,FH08,FK09} and also agree with results from numerical coagulation
simulations that model the growth of KBOs carried out by other groups
\citep[e.g.][]{KL99,K02}.

\subsection{Intermediate sized bodies: Velocity Dispersion and Growth}\label{inter}
So far we have only considered two sizes of bodies, small ones and large
ones. We now turn our attention to intermediate size bodies with radii $R'$,
mass surface density $\Sigma'$ and velocity dispersion $v'$.

In sections \ref{superHill} and \ref{subHill2}, we show that $\Sigma'$ is of
the order $\Sigma$. This is because large bodies form by accreting small
bodies until their mass surface density reaches $\Sigma/\sigma \sim
\alpha^{3/4}$. From then on, the mass surface density remains roughly constant
in a given logarithmic mass interval. $\Sigma \sim \Sigma'$ because once
larger bodies form, the velocity dispersion of the now intermediate sized
bodies quickly grows to super Hill velocities (i.e. $v'>v_H$), which implies
that such intermediate sized bodies do not grow significantly on the growth
timescale of the largest KBOs and that they are not consumed efficiently by
the larger objects. The size distribution of such intermediate sized bodies
and $\Sigma'$ therefore remain constant. During the brief period over which
$v'<v_H$, intermediate sized bodies are efficiently accreted by the largest
objects, but we show in section \ref{subHill2} that this phase is very short
and that it extends over less than a factor of 2 in radius. As a result,
$\Sigma'$ is of the order of, but slightly depleted compared to, $\Sigma$.

\subsection{Intermediate sized bodies with $v'> v_H$}\label{superHill}
There are three different velocity regimes for intermediate sized bodies that
we have to consider separately. In the first regime the intermediate sized
bodies' velocity dispersion exceeds the Hill velocity of the largest bodies,
i.e. $v'>v_H$, and the bodies themselves are sufficiently small such that
their velocity dispersion is not efficiently damped by dynamical friction. In
other words, the dynamical friction timescale for these intermediate sized
bodies exceeds the growth timescale of the large bodies.  The velocity
dispersion of these bodies is therefore dominated by gravitational stirring
from the large bodies and it grows on the same timescales as the size of the
large bodies. This implies that $v' \sim u$.
\begin{equation}
\frac{v'}{v_H}\sim \frac{u}{v_H} \sim \left(\frac{\Sigma}{\sigma
  \alpha}\right)^{1/2} \sim \alpha^{-1/8}
\end{equation}
where we substituted $\Sigma/\sigma \sim \alpha^{3/4}$ from equation (\ref{e6})
in the last step. Equating the dynamical friction timescale for bodies of size
$R'$ to the growth timescale of the large bodies and solving for $R'$, we have
\begin{equation}
\frac{R'}{R} \sim \frac{\Sigma}{\sigma} \alpha^{-1/2} \sim \alpha^{1/4}.
\end{equation}
This implies that $v' \sim u$ for intermediate sized bodies with $R' \sim
\alpha^{1/4} R \sim 0.1 R$ and smaller. For bodies above this size, damping by
dynamical friction is important.

In the second velocity regime, which applies for bodies larger than
$\alpha^{1/4} R \sim 0.1 R$ and that have $v'>v_H$, the evolution of the
velocity dispersion is dominated by gravitational stirring from the large
bodies and damping by dynamical friction generated by the small bodies. This
yields a velocity dispersion that is governed by
\begin{equation}\label{e9}
\frac{1}{v'} \frac{dv'}{dt} \sim \Omega \frac{\Sigma}{\rho R} \alpha^{-2}
\left(\frac{v'}{v_H}\right)^{-4} -
\Omega \frac{\sigma}{\rho R'} \alpha^{-2} \left(\frac{u}{v_{H}'} \right)^{-4}.
\end{equation}
Balancing the stirring and damping rates and substituting $v_H'= v_H (R'/R)$,
we find that $v'$ is given by
\begin{equation}\label{e10}
\frac{v'}{v_H} \sim \alpha^{-1/2}\left( \frac{\Sigma}{\sigma}
\right)^{3/4} \left( \frac{R'}{R}
\right)^{-3/4} \sim \alpha^{1/16} \left(\frac{R'}{R} \right)^{-3/4}
\end{equation} 
where we substituted for $u$ from equation (\ref{e2}) and used $\Sigma/\sigma
\sim \alpha^{3/4} $ in the last step.  Equation (\ref{e10}) yields that $v'
\sim v_H$ for KBOs with $R' \sim \alpha^{1/12} R \sim 0.5~R$. Therefore, $v'>
v_H$ for bodies with $R'\lesssim 0.5 R$. This implies that KBOs with radii
smaller than $\sim 0.5 R$ have super-Hill velocities (i.e., $v'> v_H$) and
those with radii larger than $\sim 0.5 R$ will have sub-Hill velocities
(i.e., $v'<v_H$).

The growth of intermediate sized bodies that have $v'>v_H$ and
$R'> \alpha^{1/4} R \sim 0.1 R$ is given by
\begin{equation}\label{e11}
\frac{1}{R'} \frac {dR'}{dt} \sim \Omega \frac{\sigma}{\rho R}
\left(\frac{R'}{R}\right) \left(\frac{u}{v_H}\right)^{-2} \alpha^{-1} + \Omega
\frac{\Sigma'}{\rho R}\left(\frac{R'}{R}\right)
\left(\frac{v'}{v_H}\right)^{-2} \alpha^{-1}.
\end{equation}
Substituting for $u$ from equation (\ref{e2}) and using again the result that
$\Sigma/\sigma \sim \alpha^{3/4}$ we can write the above expression as
\begin{equation}\label{e12}
\frac{1}{R'} \frac {dR'}{dt} \sim \Omega \frac{\Sigma}{\rho R}
\left(\frac{R'}{R}\right) \alpha^{-3/2} + \Omega
\frac{\Sigma'}{\rho R}\left(\frac{R'}{R}\right)
\left(\frac{v'}{v_H}\right)^{-2} \alpha^{-1}
\end{equation}
where the first term corresponds to growth by the accretion of small bodies
and the second term to growth by mergers of comparable sized bodies. Since $v'
> v_H$, we find when comparing the magnitude of the two growth terms in
equation (\ref{e12}), that the growth is dominated by the accretion of small
bodies. We assumed that $\Sigma'$ is of the order of $\Sigma$, which we
confirm in section \ref{subHill2}. Intermediate sized bodies with $R' \lesssim
0.5 R$ therefore grow predominantly by accreting small bodies and their growth
rate is reduced by a factor of $(R'/R)$ compared to bodies of size $R$. This
implies that bodies of size $R'$ do not get the chance to grow, compared to
the growth time scale of bodies of size $R$.

These intermediate sized bodies do not contribute significantly to the growth
of the large bodies. When examining the contributions from bodies with
$v'>v_H$ to the growth of bodies of size $R$, we have
\begin{equation}\label{e41}
\frac{1}{R} \frac {dR}{dt} \sim \Omega \frac{\Sigma}{\rho R} \alpha^{-3/2} +
\Omega \frac{\Sigma'}{\rho R} \left(\frac{v'}{v_H}\right)^{-2} \alpha^{-1}
\end{equation}
where the first and second term correspond to growth by merging with bodies of
size $R$ and $R'$, respectively. Since the first term in equation (\ref{e41})
exceeds the second, large bodies grow predominantly by accreting small bodies
and large bodies. Intermediate sized bodies with $v'>v_H$ are not important for
this growth. Furthermore, such intermediate sized bodies are only
inefficiently accreted by bodies of size $R$. This is apparent when examining
the rate of change of their surface density due to accretion onto bodies of
size $R$:
\begin{equation}\label{e42}
\frac{1}{\Sigma'} \frac {d\Sigma'}{dt} \sim \Omega \frac{\Sigma}{\rho R}
\left(\frac{v'}{v_H}\right)^{-2} \alpha^{-1}.
\end{equation}
Comparing equation (\ref{e42}) with equation (\ref{e4}), we find that
$\Sigma'$ does not change significantly on the growth timescale of the large
bodies, which implies that their surface density is not altered due to
accretion onto large bodies. As a result the mass surface density per
logarithmic mass interval of intermediate sized bodies with $v'>v_H$ remains
constant in time.

\subsection{Intermediate sized bodies with $v'< v_H$}\label{subHill2}
Intermediate sized bodies that are about half the size of the large
bodies and larger will have a velocity dispersion that is smaller than
the large bodies' Hill velocity. The evolution of the velocity dispersion of
such intermediate sized bodies is determined by gravitational stirring from
large bodies and damping by dynamical friction generated by the small
bodies. The expression for the evolution of $v'$ can be written as
\begin{equation}\label{e13}
\frac{1}{v'} \frac{dv'}{dt} \sim \Omega \frac{\Sigma}{\rho R} \alpha^{-2}
\left(\frac{v'}{v_H}\right)^{-1} - \Omega \frac{\sigma}{\rho R'} \alpha^{-2}
\left(\frac{u}{v_{H}'} \right)^{-4}.
\end{equation}
Balancing the stirring and damping rates and substituting $v_H'= v_H (R'/R)$,
we can write $v'$ as
\begin{equation}\label{e14}
\frac{v'}{v_H} \sim \alpha^{-2}\left( \frac{\Sigma}{\sigma}
\right)^{3} \left( \frac{R'}{R}
\right)^{-3} \sim \alpha^{1/4} \left(\frac{R'}{R} \right)^{-3}
\end{equation} 
where we substituted for $u$ from equation (\ref{e2}) and used $\Sigma/\sigma
\sim \alpha^{3/4} $ in the last step. Strictly speaking, we should multiply
the right hand-side of equation (\ref{e14}) by a logarithmic factor, which is
given by $3 \times Log(R/R_{*})$, where $R_{*}$ is equal to $R'$, if
$v'<v_H'$. If $v'>v_H'$ then $R_*$ is the radius that corresponds to bodies
with a velocity dispersion that is less than or equal to $v'_H$. The base of
the logarithm is the same as the base of the logarithmic mass interval over
which $\Sigma$ is defined. This logarithmic factor arises because all bodies
larger than or equal to $R'$ that have a velocity dispersion less than $v_H'$
contribute to the gravitational stirring for $q=4$. This is a key difference
compared to the super-Hill velocity regime where the stirring is dominated by
the large bodies alone.

The growth of intermediate sized bodies with $R' > \alpha^{1/12} R \sim 0.5
R$ is given by
\begin{equation}\label{e15}
\frac{1}{R'} \frac {dR'}{dt} \sim \Omega \frac{\sigma}{\rho R}
\left(\frac{R'}{R} \right) \left(\frac{u}{v_H}\right)^{-2} \alpha^{-1} +
\Omega \frac{\Sigma'}{\rho R'} F_{acc}
\end{equation}  
where 
\begin{equation}\label{e16}
F_{acc} = \left\{ \begin{array}{ll} \alpha^{-1}
\left(\frac{v'}{v_H'}\right)^{-2} & \textrm{if $v' > v_H'$}\\ 
\alpha^{-3/2} & \textrm{if $v' < v_H'$.}\\
\end{array} \right.
\end{equation}
The radius, $R_{v_H'}$, for which $v' \sim v_H'$ can be found from equation
(\ref{e14}) when including the logarithmic factor and it is given by
$R_{v_H'}/R \sim \alpha^{1/16} (3 \times Log[R/R_{v_H'}])^{1/4}$. The
$v'>v_H'$ regime given in equation (\ref{e15}) yields the same expression for
the accretion rate as given by equation (\ref{e12}). The growth in this case,
as shown above, is dominated by the accretion of small bodies. In the second
case, which $v' < v_H'$, the rate of growth of bodies with radii $R'$ is
\begin{equation}\label{e18}
\frac{1}{R'} \frac {dR'}{dt} \sim \Omega \frac{\Sigma}{\rho R}
\left(\frac{R'}{R}\right) \alpha^{-3/2} + \Omega
\frac{\Sigma'}{\rho R'}\alpha^{-3/2}
\end{equation}
where we substituted again for $u$ from equation \ref{e2} and used
$\Sigma/\sigma \sim \alpha^{3/4}$. The first term corresponds to growth by the
accretion of small bodies and the second to growth by merger of comparable
sized bodies. Comparing the magnitude of the two growth terms in equation
(\ref{e18}) we find that, unlike in the regime discussed above, the growth of
bodies with radius $R'$ is dominated by the accretion of similar sized bodies
rather than by the accretion of small bodies, again assuming that $\Sigma'$ is
of the order of $\Sigma$, which we show below. We note here that similar to
the sub-Hill velocity excitation rate, the mass accretion rate that
corresponds to the accretion of comparable sized bodies, should have been
multiplied by a logarithmic factor, which is given by $3 \times
Log[R'/R_{v_H'}]$. The maximum value for $R'/R_{v_H'}$ is given by $
\alpha^{-1/12} \sim 2$. Strictly speaking, we should also have included this
logarithmic factor in the derivation of equation (\ref{e6}). However, since
this logarithmic factor only provides a small correction it does not warrant
to be included here, since we have been neglecting factors of order unity
throughout. Although the overall growth proceeds in the runaway regime (i.e,
$(\rm{d Log}[R]/dt)/(d \rm{Log}[R']/dt)=R/R'$), large bodies with $v'<v_H'$
grow in an orderly fashion with respect to each other if $q=4$, which implies
that their radii converge. This leads to a steepening in the size distribution
at very large radii, but only for bodies with radii that have a corresponding
velocity dispersion such that $v'<v_H'$, which corresponds to
less than a factor of 2 in radius. The largest effect this can have is to
reduce $\Sigma$ by a factor of $\sim 2$, such that $\Sigma' \sim 0.5 \Sigma$,
since orderly growth will cease when $\Sigma/R \sim \Sigma'/R'$.

Intermediate sized bodies with $v'<v_H$ significantly contribute to the
growth of large bodies, since
\begin{equation}
\frac{1}{R} \frac {dR}{dt} \sim \Omega \frac{\Sigma}{\rho R} \alpha^{-3/2} +
\Omega \frac{\Sigma'}{\rho R} \alpha^{-3/2}
\end{equation}
where the first and second term correspond to growth by merging with bodies of
size $R$ and $R'$, respectively. This implies that for $q=4$, i.e. $\Sigma
\sim \Sigma'$, large bodies of size $R$ grow at comparable rates by accreting
bodies of their own size and bodies of size $R' \gtrsim 0.5 R$. Because these
intermediate sized bodies are efficiently accreted by large bodies,
their surface density is altered at a rate
\begin{equation}\label{e43}
\frac{1}{\Sigma'} \frac {d \Sigma'}{dt} \sim \Omega \frac{\Sigma}{\rho R}
\alpha^{-3/2}.
\end{equation}
Comparing equation (\ref{e43}) with equation (\ref{e4}), we find that
$\Sigma'$ changes on the order of the growth timescale of the large bodies. As
a result, the mass surface density per logarithmic mass interval of
intermediate sized bodies with $v'<v_H$ gets depleted, because such bodies
are efficiently consumed by large bodies. However, since
the size range is very small, such bodies are only depleted by factor of a few
before their velocity dispersions become super-Hill at which point they are no
longer efficiently accreted. The orderly growth of bodies with $v'<v_H'$
relative to each other and the efficient consumption of bodies with $v'<v_H$
are responsible for the steepening of the KBO size distribution seen at
$R'>R_{v'_H}$ in Figure \ref{fig2}.

\section{Coagulation Simulations}\label{s2}
\subsection{Model}
The aim of our coagulation simulation is to test our analytic results outlined
above. Its purpose is to capture the dominating physical processes that give
rise to the KBO size distribution. We attempt by no means to present the most
detailed or precise KBO formation simulation, since several such works already
exist in the current literature \citep[e.g.][]{KL99,K02,KB04}. We therefore
neglect factors of order unity in the accretion, stirring and damping
rates. In addition, we neglect effects of gas damping and possible dynamical
stirring from Neptune. We investigate the KBO growth in a single annulus
located at roughly 40~AU from the Sun with a width of about 10~AU. Most of our
simulations start with a total mass of about 40 Earth masses in small
planetesimals, which corresponds to $\sigma \sim 0.65~\rm{g~cm^{-2}}$. This
mass surface density is consistent with extrapolations of the MMSN \citep{H81}
to 40~AU, after it has been enhanced about 6 fold as required for the
formation of Uranus and Neptune \citep[e.g.][]{GLS042,DB10}. We follow the
mass growth and the evolution of the velocity dispersion of the KBOs using
Safronov's statistical approach \citep{S72}. Since we are primarily concerned
with the initial growth phase, we assume here that all collisions lead to
accretion. This assumption is justified during the initial KBO growth and
remains justified even once the planetesimal's velocity dispersion has been
excited above its own escape velocity as long as the growth timescale of the
large KBOs is short compared to the planetesimal-planetesimal collision
time. This condition is generally fulfilled, if the initial planetesimals are
of the order of a kilometer in size or larger, since, for 1-km-sized bodies,
the planetesimal-planetesimal collision time is comparable to the formation
time of Pluto. We therefore neglect the effect of destructive planetesimal
collisions here, but we discuss how they would effect our results, had they
been important, in section 4.

Below we give the relevant stirring, damping and accretion rates that
determine the growth and velocity evolution in the Kuiper belt. A detailed
derivation of the these rates can, for example, be found in \citet{GLS04}. The
accretion rate between bodies of two different mass bins is given by
\begin{equation}
\mathcal{R}_{\rm{coll}} \sim \Omega \frac{N_B \Sigma_{s}}{\rho R_B}
  \left(\frac{M_B}{M_S}\right) \times \left\{ \begin{array}{ll} 1 &
  \textrm{$v_{esc_B} < v_{rel}$}\\ \alpha^{-1}
  \left(\frac{v_{rel}}{v_{H_B}}\right)^{-2}& \textrm{$v_{H_B} < v_{rel} <
  v_{esc_B}$}\\ \alpha^{-3/2} & \textrm{$v_{rel} < v_{H_B}$}\\
  \end{array} \right.
\end{equation}
where the subscripts `s' and `B' correspond to the mass bin with the smaller
and larger bodies, respectively. The number of bodies in a given mass bin is
given by $N$ and $v_{rel}$ is the relative velocity, which was approximated by
$v_{rel}=Max[v_s,v_B]$. Note, we quote here the sub-Hill accretion rate
applicable to large objects, since the velocity dispersion of the small bodies
is super-Hill throughout the growth.

The corresponding velocity evolution for each mass bin is dominated by the
following processes. The velocity dispersion of the mass bin that corresponds
to the smaller objects is viscously stirred by the object in the larger
mass bin at a rate given by
\begin{equation}
\frac{1}{v_s}\frac{dv_s}{dt}\sim \Omega \frac{\Sigma_{B}}{\rho R_B} \times
\left\{
\begin{array}{ll} 1  & \textrm{$v_{esc_B} < v_{rel}$}\\ \alpha^{-2}
\left(\frac{v_{rel}}{v_{H_B}}\right)^{-4} & \textrm{$v_{H_B} < v_{rel} <
v_{esc_B}$}\\ \alpha^{-2} \left(\frac{v_{rel}}{v_{H_B}}\right)^{-1} & \textrm{$v_{rel} < v_{H_B}$}\\
\end{array} \right.
\end{equation}
The velocity dispersion of the bodies in the larger of the two mass bin is in
turn damped by the smaller bodies. The damping rate of the large bodies
velocity dispersion is given by
\begin{equation}
\frac{1}{v_B}\frac{dv_B}{dt} \sim -\Omega \frac{\Sigma_{s}}{\rho R_B} \times
  \left\{ \begin{array}{ll} 1 & \textrm{$v_{esc_B} < v_{rel}$}\\ \alpha^{-2}
  \left(\frac{v_{rel}}{v_{H_B}}\right)^{-4} & \textrm{$v_{H_B} < v_{rel} <
  v_{esc_B}$}\\ \alpha^{-2} & \textrm{$v_{rel} < v_{H_B}$}\\
\end{array} \right.
\end{equation}
We implemented the above equations in our coagulation code and followed the
growth and velocity evolution of the various mass bins.

\subsection{Results}
\subsubsection{KBO Growth and Velocity Evolution}
\begin{figure}[htp]
\centerline{
\epsfig{file=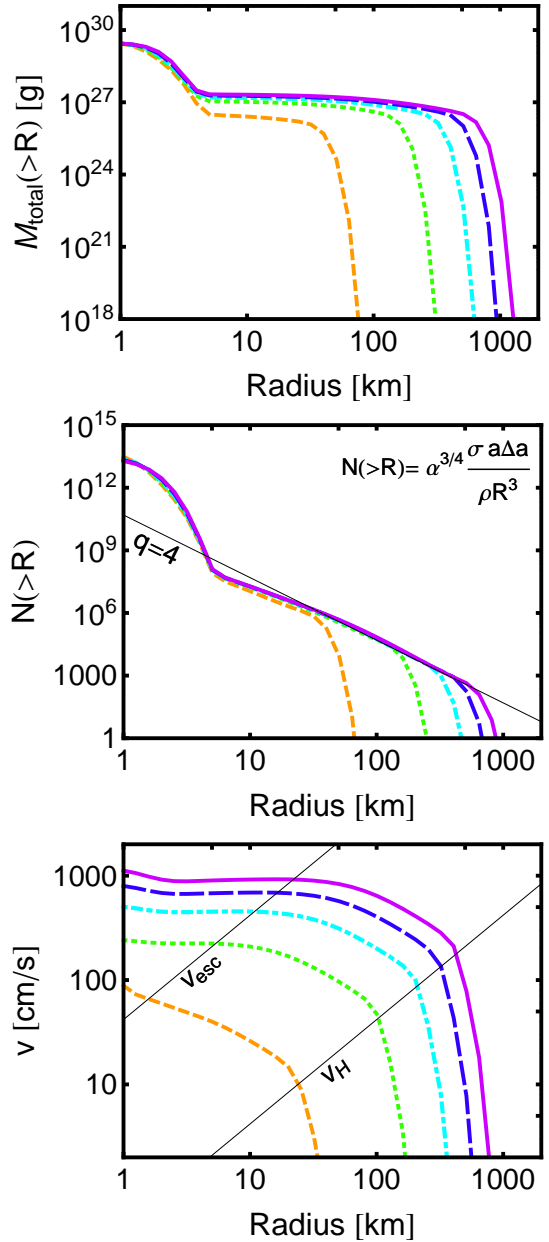, scale=0.73}}
\caption{Evolution of the cumulative mass distribution, the cumulative number
distribution and the velocity dispersion during KBO growth as a function of
time at $3 \times 10^7$ years (dashed orange line), $4 \times 10^7$ years
(dotted green line), $5 \times 10^7$ years (dot-dashed turquoise line), $6
\times 10^7$ years (long-dashed blue line), and $7 \times 10^7$ years (solid
purple line). The growth of the large KBOs ($R \gtrsim 50~\rm{km}$) becomes
self-similar from $4 \times 10^7$ years onwards, i.e. the shapes of the
size-distribution and of the velocity-distribution remain unchanged but the
KBO size and velocity continue to grow with time. The slope of the large KBO
size distribution is $q \sim 4$. A power-law index of $q=4$ corresponds to a
horizontal line in the top panel of this figure. The thin black
line in the middle panel of this figure represents the KBO size distribution
with $q=4$, as predicted by our analytic theory and its expression is given in
the top righthand corner. The bottom panel shows the evolution of velocity
dispersion during KBO growth. The escape velocity and Hill velocity (assuming
a KBO material density of $1~\rm{g/cm^3}$) are given as a function of size by
the upper and lower thin black line, respectively. We refer the reader to
Figure \ref{fig3} for a detailed comparison between our analytic theory and
the velocity evolution found from our simulations. Our analytic theory and the
numerical coagulation results are in excellent agreement.}
\label{fig1}
\end{figure}

\begin{figure} [htp]
\centerline{\epsfig{file=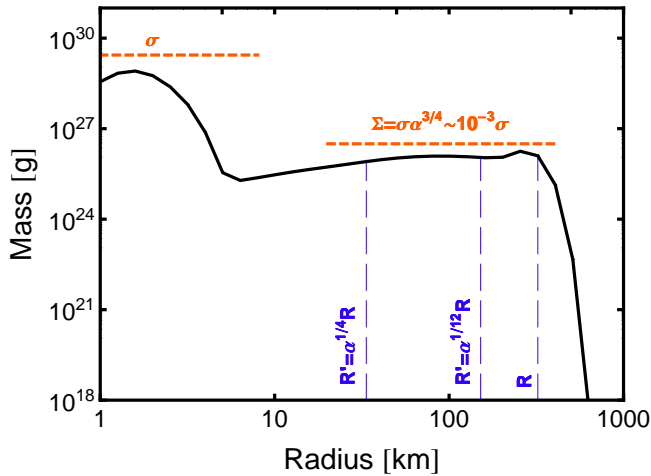, scale=0.76}}
\caption{KBO size distribution as a function of radius at $5 \times
  10^{7}~\rm{years}$ from the same coagulation simulations as shown in Figure
  \ref{fig1}. The y-axis corresponds to the mass in a given $\rm{Log_{2}}$
  mass interval. The total mass in small planetesimals is given by $\sigma$
  and its values is given by the orange dashed line drawn below its
  symbol. Similarly, the mass in large KBOs in a given $\rm{Log_2}$ mass
  interval is denoted by $\Sigma$ and its predicted value from section
  \ref{s1} is given by orange dashed lines drawn below it. The value for
  $\Sigma$ from our analytic work and numerical coagulation simulation agree
  within a factor of 2. This agreement can even be improved, if we account for
  the fact that all logarithmic mass bins with $v'<v_H$ contribute to the
  growth of the largest bodies about equally (see discussion following
  equation (\ref{e18}) in section \ref{s1}), which would reduce the predicted
  value of $\Sigma$ somewhat. The dashed vertical lines mark
  the radii separating the different velocity dispersion regimes, as discussed
  in section \ref{s1}.}
\label{fig2} 
\end{figure}

\begin{figure} [htp]
\centerline{\epsfig{file=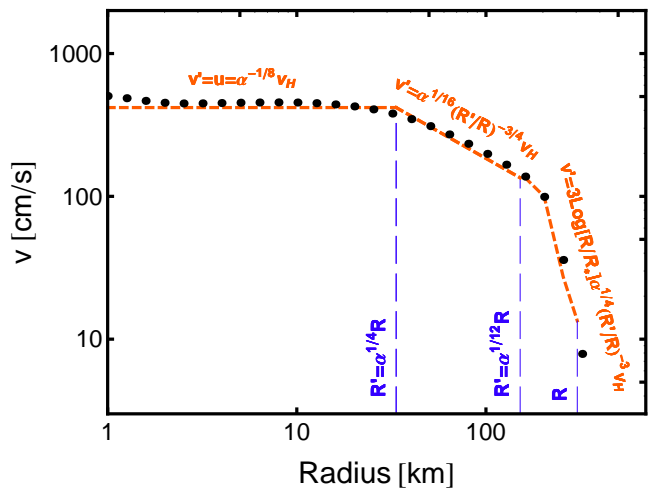, scale=0.76}}
\caption{Velocity dispersion as a function of radius at $5 \times
  10^{7}~\rm{years}$. The results from our coagulation simulation are given by
  the black points, the dashed orange lines are our analytic predictions for
  the velocity dispersion with their equations given above each segment. The
  dashed vertical lines mark the radii separating the different velocity
  dispersion regimes, as discussed in section \ref{s1}.}
\label{fig3} 
\end{figure}

As initial conditions we start with an equivalent of about 40 Earth masses, in
1-km sized planetesimals with an initial velocity dispersion of 3 times their
Hill velocity. We follow the KBO growth and velocity evolution of the
different mass bins for 70~Myrs. The results of our coagulation simulation are
shown in Figures \ref{fig1}, \ref{fig2}, and \ref{fig3}. 

Figure \ref{fig1} shows the evolution of the cumulative mass distribution, the
cumulative number distribution and the velocity dispersion during KBO growth
as a function of time. It is apparent from Figure \ref{fig1} that the growth
and velocity evolution become self-similar once $\Sigma \sim \alpha^{3/4}
\sigma \sim 10^{-3} \sigma$, i.e. the shape of the size-distribution and
velocity-distribution remains unchanged, while the maximum KBO size continues
to grow.

The two upper panels of Figure \ref{fig1} and Figure \ref{fig2} show that the
size-distribution of large KBOs ($R \gtrsim 50~\rm{km}$) indeed follows a
power-law with $q=4$ as predicted by our analytic treatment in section
\ref{s1}. This implies a roughly equal amount of mass per logarithmic mass bin
for large KBOs. Moreover, the middle panel of Figure \ref{fig1} and Figure
\ref{fig2} show that the mass ratio of large to small KBOs,
i.e. $\Sigma/\sigma$, found in our coagulation simulation agrees very well
with our analytic prediction that $\Sigma/\sigma \sim \alpha^{-3/4} \sim
10^{-3}$. The simulations confirm our analytic results and suggest that the
total mass in large objects that we see in the Kuiper belt is not arbitrary
but an outcome of the KBO growth and that it is roughly $10^{-3}$ of the
initial planetesimals mass. This result is in excellent agreement with the
actual observed mass in large KBOs and formation from a MMSN type disk. Our
work therefore suggests that the Kuiper belt did not contain two to three
order of magnitude more mass in large KBOs as has been proposed by some
models \citep{W02,TGM05}.
 
Figure \ref{fig3} shows a comparison of the velocity dispersion from our
coagulation simulation and our analytic results derived in section \ref{s1}. It
displays very good agreement between the velocity dispersion that we derived
analytically for the various size regimes, and the results from our
coagulation simulation.

Figures \ref{fig1}-\ref{fig3} show that our analytic work captures the
essential features of KBO growth and that analytic theory and the numerical
coagulation results are in excellent agreement. We are able to successfully
explain the slope ($q=4$) and amplitude ($\Sigma \sim \alpha^{-3/4} \sigma$)
of the large KBO size distribution and the evolution of the velocity
dispersion in the various velocity and size regimes.

\subsubsection{Initial Planetesimal size distribution and sizes}
\begin{figure} [htp]
\centerline{\epsfig{file=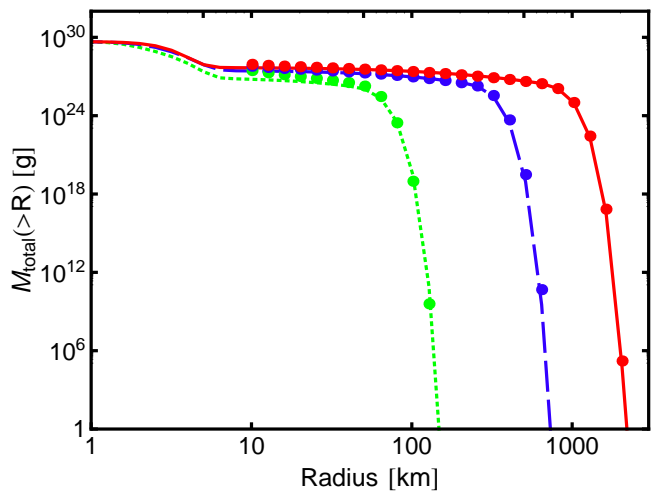, scale=0.76}}
\caption{Direct comparison of KBO growth from two different initial
  conditions. The lines correspond to starting with only 1-km-sized
  bodies. The points represent the KBO growth that resulted when starting with
  the same total mass as before, but in addition to having most of the mass in
  1-km-sized objects we added a second population of 10-km-sized KBOs that
  contain $10^{-3}$ of the total mass. The cumulative mass distribution of the
  small objects is omitted here, since it is the size-distribution of large
  KBOs that we want to compare here. The agreement between the two simulations
  is striking. The shape, amplitude (i.e. the ratio of $\Sigma/\sigma$) and
  power-law slope is therfore not a result of the initial conditions but
  represents an `equilibrium state' that the system evolves to. The dotted
  green line corresponds to a time of about $3 \times 10^7$ years since the
  start of the simulation, whereas the green points correspond to only about
  $10^{6}$ years since the start of the simulation. Due to the different
  initial conditions it took a different amount of time to grow to 50-km-sized
  KBOs, but the growth timescale once these 50-km-sized KBOs have formed
  (green dashed line and points) becomes the same. In other words, it takes
  the same amount of time to grow from 50-km-sized KBOs to Pluto-sized objects
  irrespective of the initial conditions, once 50-km-sized KBOs have
  formed. The blue dashed line and points and the red solid line and points
  correspond to $1.5 \times 10^7$ years and to $5 \times 10^7$ years, since the
  formation of objects that are represented by the green dashed lines and
  points, respectively.}
\label{fig4} 
\end{figure}

We performed an additional set of coagulation simulations with different
initial conditions. In the first set we started with most of the mass in small
1-km-sized objects, just as before, but added a second population of larger
10-km-sized KBOs that contained $10^{-3}$ of the total mass. All bodies were
started with a velocity dispersion equal to 3 times their Hill velocity and we
followed their growth and velocity evolution. Figure \ref{fig4} shows the
result of this growth (points) and a comparison with the cumulative mass
distribution at various times for KBOs that grew from a single population of
1-km sized planetesimals (lines). The similarity between the two distributions
is striking. The same power-law for the large KBO size distribution emerges
and the mass ratio in large and small KBOs becomes also the same in both
simulations. These results highlight the power of the `two groups
approximation' that we used to derive the analytic results in section \ref{s1}
and validate our assertion that the growth of large KBOs develops towards a
state where small and large bodies contribute about equally to the growth. The
overall growth timescale to reach Pluto-mass objects differs in these two
simulations. Starting from 40 Earth masses in small planetesimals, it takes in
our simulations about 70-80~Myrs to form Pluto-sized objects if initially all
KBOs are 1-km in radius but only 40-50~Myrs if, in addition to the
1-km-sized objects, there also existed a small population of 10-km-sized
KBOs. The difference in the growth time results from the timescale it takes
1-km-sized objects to grow into 10-km sized bodies, if none are already
present. When only comparing the growth timescale from 50-km-sized objects to
1000-km sized objects, we find that they are the same in the two scenarios
(see figure \ref{fig4}), confirming that it is the initial growth of objects
to tens of kilometers in size that give rise to the overall difference in
Pluto-formation time between the two simulations. The final shape of the KBO
size-distribution and the total mass in large KBOs (as a fraction of the
initial mass) is therefore independent of the initial size distribution of the
planetesimals. We also find the same KBO size distribution and overall mass in
large KBOs, if the initial planetesimals have a power-law size distribution
with various power-law indexes and radii ranging from 1 to 10-km. In addition,
we confirm that our results are independent of the details of the initial
planetesimal velocity dispersion, as long as it is below their escape velocity.

\begin{figure} [htp]
\centerline{\epsfig{file=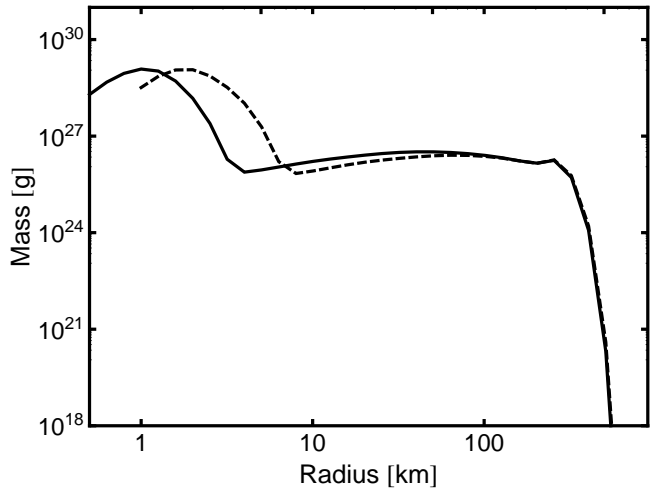, scale=0.76}}
\caption{Comparison of KBO growth from two different initial planetesimal
  sizes. In each simulation, all planetesimals started out with one size. The
  solid and dashed line correspond to an initial planetesimal size of 500~m
  and 1~km, respectively. The mass distribution is shown after a growth time
  of $\sim 5 \times 10^{7}~\rm{years}$. The y-axis corresponds to the mass in
  a given $\rm{Log_{2}}$ mass interval. The shape of the size distribution
  does not depend on the initial planetesimal size, as long as the
  planetesimal-planetesimal collision timescale exceeds the KBO growth
  timescale, such that dynamical cooling and possible mass loss due to
  frequent planetesimal collisions can be neglected.}
\label{fig5}
\end{figure}

We also performed a second set of coagulation simulations, which we started
with the same total planetesimal mass but with an initial planetesimal size of
500~m in radius rather than 1~km. Figure \ref{fig5} shows that the shape of
the size distribution does not depend on the initial planetesimal size. This
result remains valid as long as the initial planetesimal size is large enough
such that the planetesimal collision timescale exceeds the KBO
growth timescale.

These results confirm earlier findings by \citet{KL99}, who note that the
power-law slope for large objects is remarkably independent of the input
parameters and initial conditions. Our work offers an explanation of why this
is so. Since the size distribution of large KBOs reaches an equilibrium state
that evolves self-similarly, the signature of the initial planetesimal
size-distribution and their initial velocity dispersion are erased. The slope
of the KBO size distribution and the formation timescales that we find in our
simulations for large KBOs is in agreement with results from previous
coagulation simulations \citep{KL99,K02,KB04}.

\subsubsection{Different values of $\alpha$}
\begin{figure} [htp]
\centerline{\epsfig{file=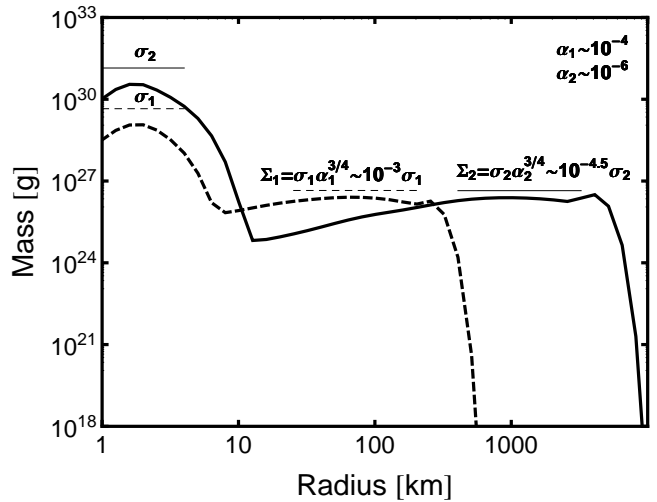, scale=0.76}}
\caption{Comparison of our coagulation results for two different values of
  $\alpha$. The dashed lines correspond to $\alpha_1 = 10^{-4}$, which is
  roughly the value for the Kuiper belt, and the solid lines correspond to
  $\alpha_2 = 10^{-6}$, which could, for example, represent a debris disk
  around a solar-mass star with a semi-major axis about 100 times larger than
  the Kuiper belt. We increased the initial planetesimal mass in the $\alpha_2
  = 10^{-6}$ simulation by a factor of $10^{1.5}$ to speed up the planetesimal
  growth. The y-axis corresponds to the mass in a given $\rm{Log_{2}}$ mass
  interval; $\sigma_1$ and $\sigma_2$ are the sum of the total mass in small
  planetesimals. $\Sigma_1$ and $\Sigma_2$ correspond to the mass in large
  KBOs in a given $\rm{Log_2}$ mass interval and their predicted values from
  section \ref{s1} are given by the lines drawn below their respective
  symbols. The agreement between theory and simulation for both systems is
  clearly shown in Figure \ref{fig6}.}
\label{fig6} 
\end{figure}

Since our analytic result for the mass ratio between large and small KBOs
(i.e., $\Sigma/\sigma \sim \alpha^{3/4}$) can be expressed as a function of
$\alpha$ alone, we performed an additional set of coagulation simulations for
a value of $\alpha$ that was decreased by a factor of 100. Figure \ref{fig6}
shows the coagulation results for $\alpha=10^{-4}$ (dashed lines), which is
roughly the value for the Kuiper belt, and $\alpha=10^{-6}$ (solid lines). We
increased the initial planetesimal mass in the $\alpha = 10^{-6}$ simulation
by a factor of $10^{1.5}$ to speed up the planetesimal growth.  Due to this
mass increase, the resulting mass per logarithmic mass interval in large
bodies should be the same in both simulations. This is indeed what we see in
Figure \ref{fig6}, confirming our analytic expression for $\Sigma$. Figure
\ref{fig6} shows also that the equilibrium growth state, where growth by
accretion of small bodies is comparable to the growth due to similar sized
mergers, is reached later for $\alpha=10^{-6}$ compared to $\alpha=10^{-4}$. We
confirmed that this is not caused by the increased planetesimal mass that was
used in the $\alpha = 10^{-6}$ simulation.

\section{Collisions And Other Considerations}\label{s3}
We have so far neglected collisions among the small bodies. In this section,
we first derive the planetesimal-planetesimal collision timescale and show
that planetesimal collisions can be neglected during the KBO growth for
planetesimal sizes of about a kilometer and larger. We then discuss how our
results would be altered, if planetesimal collisions had been important during
the KBO growth. In addition, we examine the effect of semi-collisional
accretion \citep{SG06,SR06} and frequent binary mergers on the growth of
KBOs.

The planetesimal-planetesimal collision time for objects with radius, $r$, and
with a velocity dispersion that has been excited above their escape velocity
is given by
\begin{equation}\label{ec1}
\tau_{Coll} \sim \Omega^{-1} \frac{\rho r}{\sigma}.
\end{equation}
Comparing equation (\ref{ec1}) to the Pluto formation timescale, which is
given by the inverse of equation (\ref{e5}) we find that
\begin{equation}
\frac{\tau_{Coll}}{\tau_{Pluto}} \sim \left(\frac{\Sigma r}{\sigma
R_{Pluto}}\right) \alpha^{-3/2},
\end{equation}
which is $\sim 1$ for 1-km-sized planetesimals and $\Sigma \sim \alpha^{3/4}
\sigma$. 1-km is about the size that planetesimals are expected to
have, if they formed by gravitational instability without dissipation of
internal angular momentum \citep{GW73,GLS04}. This implies that if KBOs grew
from a population of 1-km sized planetesimals then mutual planetesimal
collisions become important only once KBOs comparable to the size of Pluto
have formed.

If planetesimal-planetesimal collisions are important, then they would, most
likely, break-up the planetesimals and lead to a damping of the small bodies
velocity dispersion, which would therefore no longer grow as $R$, as we
assumed in section \ref{s1}. Instead, it would be set by balancing the
dynamical stirring from large bodies by collisional cooling. This yields
\begin{equation}\label{ce1}
\frac{u}{v_H} \sim \frac{\Sigma s}{\sigma R} \alpha^{-2}
\end{equation}
for $u<v_H$, where $s$ is the radius of the small bodies \citep{GLS04}. We
note here that $u<v_H$ is most likely the relevant case to consider, since $u$
is only slightly super-Hill when collisional cooling is neglected (see section
\ref{s1}). In this case, the growth of KBOs is given by
\begin{equation}\label{ce2}
\frac{1}{R} \frac {dR}{dt} \sim \Omega \frac{\sigma}{\rho R} \alpha^{-1}
\left(\frac{u}{v_H}\right)^{-1} + \Omega \frac{\Sigma}{\rho R} \alpha^{-3/2} 
\end{equation}
where we assumed that $\alpha^{1/2} v_H <u<v_H$. Substituting for $u$ from
equation (\ref{ce1}) and comparing the two growth rates in equation
(\ref{ce2}), we find that the two growth rates become equal when
\begin{equation}
\frac{\Sigma}{\sigma} \sim \left(\frac{R}{s}\right)^{1/2} \alpha^{5/4} \sim
10^{-1} \left(\frac{R/1000~\rm{km}}{s/1~\rm{cm}}\right)^{1/2}.  
\end{equation}
This implies, that in this case, the growth of large bodies is dominated by
the accretion of small bodies until more mass is converted into $\Sigma$,
compared to the case without frequent planetesimal collisions. The total mass
in large KBOs and the slope of their size distribution are therefore likely to
be different from our results in section \ref{s1} and \ref{s2}, if collisions
were important. The KBO growth would be halted, if collisions lead to the
onset of a collisional cascade and if the small collisional fragments are
efficiently lost from the Kuiper belt.

So far, we have treated the planetesimal accretion as collisionless, meaning
that collisions among planetesimals can be neglected while they are inside the
Hill sphere of a growing KBO. This assumption is valid, if KBOs formed from
km-sized planetesimals. However, if planetesimals are of the order of a meter
in size or smaller, either because they formed small, or because they were
broken into small pieces, then they are more likely to collide with each other
inside a KBO's Hill sphere than to accrete onto the growing KBO directly
\citep{SR06}. We call this semi-collisional accretion. Such planetesimal
collisions inside the Hill sphere lead to the formation of an accretion disk
around the KBOs. If KBOs grew by semi-collisional accretion than their growth
could have been very fast, because the effective radius for accretion, in this
case, is of the order of their Hill sphere. The KBO growth would be dominated
by the accretion of small planetesimals for much longer in the semi-collisional
regime, compared to the growth scenario discussed and investigated in
sections \ref{s1} and \ref{s2}.

Finally, the growth of KBOs might also have been aided by merging comparable
mass Kuiper belt binaries. As we discuss in section \ref{sec-binaries} in
detail, a significant fraction of KBOs reside in comparable mass binary
systems and such systems likely formed by dynamical processes such as by three
body capture, or by binary formation aided by dynamical friction
\citep{W02,GLS02,F04,A05,A07}. Once formed, the mutual semi-major axis of such
binary systems starts to shrink due to dynamical friction generated by the
small planetesimals \citep{GLS02}. This eventually leads to merging between
the binary components. This channel of growth could be important, because the
binary formation time and the time it take for a binary to spiral in until
contact occurs are comparable to the timescale on which large KBOs grow
\citep{GLS02}. If merging of binary components is important in the overall KBO
growth then the growth rate would be enhanced compared to equation
(\ref{e4}). This results in a reduced mass surface density for large KBOs, but
the power-law slope of the large KBO size distribution will remain unchanged.

\section{Implications for Kuiper Belt Binaries}\label{sec-binaries}
Our coagulation results also have interesting implications for the formation
of Kuiper belt binaries. A significant fraction of KBOs are part of a binary
system. The binary fraction varies for different dynamical classes and it is
highest in the cold classical belt \citep{NGS08}, where it is about 30\%. More
than 70 binaries have been discovered in the Kuiper belt to date and their
number continues to rise. High-mass ratio binary systems, including
Pluto/Charon, likely formed in a collision and subsequent tidal
evolution. However, the majority of Kuiper belt binaries consist of comparable
mass companions with wide separations and have too much angular momenta to
have formed by the same mechanism. Instead these systems most likely have a
dynamical origin like, for example, binary formation by three body capture, or
binary formation aided by dynamical friction
\citep{W02,GLS02,F04,A05,A07}. The Hill sphere, the region interior to an
object's Hill radius, sets the maximum phase space available for binary
formation by such dynamical processes. Dynamical binary formation scenarios
take advantage of the increased size of the Hill radius, which is more than an
order of magnitude larger in the Kuiper belt than for similar sized objects in
the Asteroid belt and therefore make the Kuiper belt the ideal place for
the formation of wide, comparable mass binaries.

\begin{figure} [htp]
\centerline{\epsfig{file=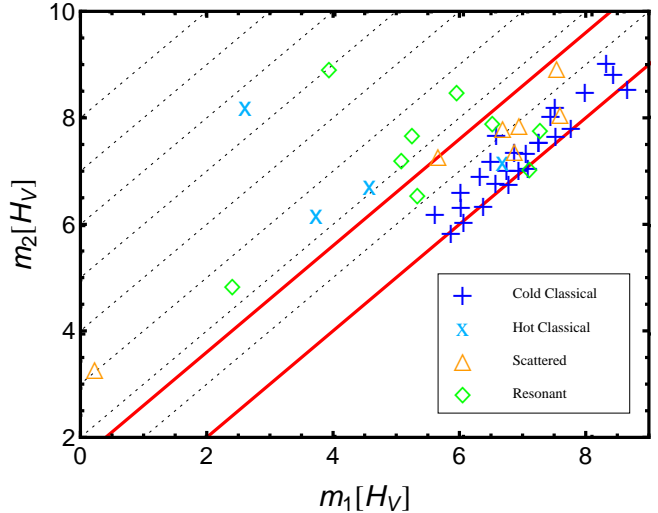, scale=0.76}}
\caption{The $H_V$ magnitudes of both components of Kuiper belt binaries,
  labeled by their respective dynamical class according to the classification
  by \citet{GMV08}. The black dotted lines, correspond to increments of 1
  magnitude difference between the binary components. The region enclosed by
  the solid red lines ranges from equal magnitude binary components to a
  maximum magnitude difference of $\Delta_{mag} \sim 1.7$. This corresponds to
  the maximum size difference that Kuiper belt binary components with similar
  albedos can have as predicted by this work, if binary formation preceded by
  dynamical processes (see section \ref{sec-binaries} for details). All
  binaries in the cold classical belt lie in the region enclosed by the solid
  red lines.  In contrast to the cold classical population, the binary
  components of other dynamical classes span a wider range in mass ratios. The
  plotted binary data was obtained from the following references: \citet{N08}
  and references therein, \citet{NGS08} and \citet{LK10}. We calculated the
  $H_V$ magnitude of each binary component from the magnitude difference
  between the primary and secondary and from their combined $H_V$ magnitude as
  given in the references above.  Pluto and Eris, which are both not part of
  the classical belt, are not shown on this plot, since they have $H_V$
  magnitudes brighter than 0 and fall therefor to the left of this figure.}
\label{fig8} 
\end{figure}

Dynamical binary formation scenarios require $v'<v_H$ for efficient formation,
since binary formation rates quickly exceed the age of the solar system, once
the velocity dispersion significantly exceeds the Hill velocity
\citep{N08,SR07}. Since KBO growth and binary formation occurs concurrently,
we can predict the maximum mass ratio for such binary systems because we know
the range of KBO radii that have sub-Hill velocities at any given stage during
the growth. From equation (\ref{e10}) we have $v'<v_H$ for KBOs with radii
greater than $R' \sim \alpha^{1/12} R \sim 0.5~R$. This implies a maximum mass
ratio of $ \sim \alpha^{-1/4} \sim 10$ between the primary and secondary and
translates into a maximum magnitude difference between the binary components,
assuming similar albedo, of about $\Delta_{mag} \sim 1.7$. This provides an
explanation for the strong clustering in $\Delta_{mag}$ of binaries in the
cold classical belt. This clustering was first pointed out by \citet{N09} and
is shown in Figure \ref{fig8}. Figure \ref{fig8} shows that all binaries in
the cold classical belt have $\Delta_{mag} <1.7$, i.e. they lie in the region
enclosed by the solid red lines. Our coagulation work therefore suggests that
the observed binaries in the classical belt formed by dynamical processes
during the growth of the KBOs themselves and that this formation most likely
took place in situ. The cold classical belt is therefore not only distinct in
its inclination distribution \citep{BT01,E05,G10}, color \citep{TR00,TB02,P08}
and binary fraction \citep{SN06}, but also in terms of the type of its
binaries, i.e. they are exclusively comparable mass ratio binaries with
$\Delta_{mag}<1.7$ \citep{N08,N09}. Figure \ref{fig8} shows that, in contrast
to the cold classical population, other dynamical classes have binaries with a
wide range of mass ratios and they therefore most likely contain dynamically
and collisionally formed binary systems. The fact that the other dynamical
classes have comparable mass binaries as well as systems with a wider range of
component sizes is consistent with the idea that they are a superposition of
two original populations, namely a dynamical cold (low inclination) population
that formed close to its current location and a dynamical hot (high
inclination) population that formed closer to the Sun. Kuiper belt binaries
may therefore prove to be useful probes for entangling the two original
dynamical populations, if they existed (Murray-Clay \& Schlichting 2010,
submitted).

\section{Discussion \& Conclusions}
We carried out an analytic and numerical investigation of runaway growth of
protoplanets with a focus on the Kuiper belt. Since, the Kuiper belt is a
remnant of the primordial Solar system where planet formation never reached
completion, it contains some of the least processed bodies. As a result, it
provides a snapshot of earlier stages of planet formation, which were erased
elsewhere in the Solar system where planet formation proceeded to
completion. Our results for runaway growth, which we summarize below,
therefore, do not only apply to the Kuiper belt but also to planet formation
before the onset of oligarchic growth. In addition, our findings apply to
proto planetary growth in debris disks around other stars, as long as the
protoplanets' growth time is shorter than the planetesimal-planetesimal
collision time, such that dynamical cooling and possible mass loss due to
frequent planetesimal collisions can be neglected.

In this paper, we presented analytic work that describes the growth of KBOs,
the evolution of their velocity dispersion, and that provides insights into
the underlying physical processes that give rise to the KBO size
distribution. Our work successfully explains the observed slope of the KBO
size distribution as well as the total mass that is present in large KBOs
today. In addition it predicts the maximum mass-ratio of Kuiper belt binaries
that formed by dynamical processes, which explains the observed clustering in
binary companion sizes that is seen in the cold classic belt. We confirmed our
analytic results with numerical coagulation simulations.

We find that the KBO growth proceeds as follows. Initially all the mass
resides in small planetesimals and large KBOs start to form by accreting small
planetesimals. This growth continues until growth by merging with comparable
sized KBOs become comparable to growth by accreting small bodies. We show that
this condition sets in when $\Sigma/\sigma \sim \alpha^{3/4} \sim
10^{-3}$. From that time onwards, the growth and the evolution of the velocity
dispersion become self-similar and $\Sigma$ remains roughly constant, since
the increase in $\Sigma$ by the accretion of small planetesimals is balanced
by a decease due to the accretion of large bodies. We showed that this mode of
growth leads to a KBO size distribution with power-law index $q=4$. This is in
good agreement with observations of the Kuiper belt size distribution, which
is well described by a power-law with an index that is consistent with $q=4$
within $1\sigma$ \citep[e.g.][]{TJL01,BTA04,FH108,FK108}. A single albedo is
assumed for all sizes when converting from the observed magnitude distribution
to the KBO size distribution. Therefore, possible albedo variations as a
function of size could introduce a significant uncertainty in the estimate of
$q$. The best fit value for the power-law index is typically found to somewhat
exceed $q=4$. We note that this could be due to fitting the high end of the
size distribution that corresponds to the largest KBOs, which is somewhat
steeper than the rest of the distribution (see figure \ref{fig1}). This
steeper end of the KBO size distribution is due to the fact that the largest
objects did not have enough time to grow to their steady-state abundance.

If KBOs formed by coagulation from km-sized planetesimals, then there could
not have been significantly more mass in large KBOs than what is observed
today, unless the MMSN was initially enhanced by several orders of
magnitude. This result is in good agreement with the current mass in large
KBOs, since it is about $\Sigma/\sigma \sim \alpha^{3/4} \sim 10^{-3}$
of the MMSN that was enhanced by a factor of a few, as required for the
formation of Uranus and Neptune. The observed mass in large KBOs and their
size distribution therefore support the hypothesis that KBOs formed from
a MMSN type disk by coagulation from kilometer sized planetesimals. The growth
of KBOs ended in this runaway phase with $\Sigma \sim \alpha^{3/4} \sigma \sim
10^{-3} \sigma$. In the context of planet formation this phase of runaway
growth was terminated by the onset of oligarchic growth when each large body
dominates the stirring in its own feeding zone. During oligarchic growth
$\Sigma$ continues to grow until it becomes comparable to $\sigma$.

Our understanding of the growth, the value of $\Sigma/\sigma$ and the
size distribution of large bodies differs from previous works that
investigated the details of runaway growth \citep{MFF98,MHL00,MG01}. This is
mainly due to the fact that these works did not take into account the
simultaneous evolution of the velocity dispersion during the growth but
assumed either equipartition between the different mass bins
\citep{MFF98,MG01} and/or a constant velocity dispersion for all sizes
\citep{MHL00}.

Since efficient binary formation by dynamical processes can only proceed
among KBOs that have sub-Hill velocity dispersions \citep{N08,SR07} and
because we follow the evolution of the KBO velocity dispersion as well as
their growth, we can predict the maximum mass ratio for such binary
systems. Our work yields a maximum binary mass ratio of $\alpha^{-1/4} \sim
10$, or, in other words, a maximum magnitude difference between binary
components, assuming similar albedos, of $\Delta_{mag} \sim 1.7$. This
explains the clustering in $\Delta_{mag}$ of all the observed Kuiper belt
binaries that are part of the cold classic belt, which make up about half of
all known binary systems in the Kuiper belt.

There exists some tentative observational evidence that the power-law size
 distribution might be different for KBOs with inclinations greater and
 smaller than $\sim 5^{\circ}$ \citep{BTA04,FH108,FB10}. However, \citet{FK09}
 and \citet{FH10} found no conclusive evidence supporting such a
 difference. It will be interesting to see the results of future KBO surveys
 that address this question. Should there indeed exist a difference between
 various dynamical classes then this could have very interesting implications
 for the formation and collisional evolution of the Kuiper belt.

\acknowledgements{We thank Peter Goldreich for helpful comments on this
  manuscript. For HS support for this work was provided by a Canadian
  Institute for Theoretical Astrophysics Postdoctoral Fellowship and by NASA
  through Hubble Fellowship Grant \# HST-HF-51281.01-A awarded by the Space
  Telescope Science Institute, which is operated by the Association of
  Universities for Research in Astronomy, Inc., for NASA, under contact NAS
  5-26555. The research of RS is supported by ERC, IRG and HST grants.}


\end{document}